\documentclass[pre,aps,twocolumn]{revtex4}
\usepackage{amssymb}
\usepackage{amsmath}
\usepackage{bm}
\usepackage{graphicx}
\usepackage{subfigure}

\usepackage{hyperref}

\begin{document}

\title{Correlation-Compressed Direct Coupling Analysis}

\author{Chen-Yi Gao$^{1,3}$, Hai-Jun Zhou$^{1,3,5}$\footnote{Corresponding author. Email: zhouhj@itp.ac.cn}, and Erik Aurell$^{2,4}$\footnote{Corresponding author. Email: eaurell@kth.se}}

\affiliation{$^{1}$Key Laboratory of Theoretical Physics, Institute of Theoretical Physics, Chinese Academy of Sciences, Beijing 100190, China}
\affiliation{$^2$Department of Computational Biology, KTH-Royal Institute of Technology, SE-10044 Stockholm, Sweden}
\affiliation{$^3$School of Physical Sciences, University of Chinese Academy of Sciences, Beijing 100049, China}
\affiliation{$^4$Depts of Applied Physics and Computer Science, Aalto University, FIN-00076 Aalto, Finland}
\affiliation{$^5$Synergetic Innovation Center for Quantum Effects and Applications,
Hunan Normal University, Changsha, Hunan 410081, China}

\begin{abstract}
Learning Ising or Potts models from data has become an important topic in statistical physics and computational biology, with applications to predictions of structural contacts in proteins and other areas of biological data analysis. The corresponding inference problems are challenging since the normalization constant (partition function) of the Ising/Potts distributions cannot be computed efficiently on large instances. Different ways to address this issue have hence given size to a substantial methodological literature. In this paper we investigate how these methods could be used on much larger datasets than studied previously. We focus on a central aspect, that in practice these inference problems are almost always severely under-sampled, and the operational result is almost always a small set of leading (largest) predictions. We therefore explore an approach where the data is pre-filtered based on empirical  correlations, which can be computed directly even for very large problems. Inference is only used on the much smaller instance in a subsequent step of the analysis. We show that in several relevant model classes such a combined approach gives results of almost the same quality as the computationally much more demanding inference on the whole dataset. We also show that results on  whole-genome epistatic couplings that were obtained in a recent computation-intensive study can be retrieved by the new approach. The method of this paper hence opens up the possibility to learn parameters describing pair-wise dependencies in whole genomes in a computationally feasible and expedient manner.
\end{abstract}

\date{\today}

\maketitle

\section{Introduction}
\label{sec:introduction}

Inferring interactions from multiple sequence alignment (MSA) has emerged in recent years as an important new development in statistical physics and computational biology. A main model paradigm has been to use the data to infer terms in an Ising model, if the data is Boolean (Ising spins), or to infer terms in a Potts model if the data is categorical (number of data types greater than two). From a statistical point of view these are inference problems in exponential families~\cite{WainwrightJordan2008}, while from a physical point of view the approach has been called the inverse Potts/Ising problem~\cite{Roudi2009,NguyenBergZecchina} and Direct Coupling Analysis~\cite{WeigtWhite2009}. In this paper we will use the latter term and its abbreviation DCA. DCA has been used to predict residue-residue contacts in a protein 3D structure from similar (homologous) protein sequences~\cite{WeigtWhite2009,Burger2010,jones2012,Magnus2013}, reviewed in~\cite{SteinMarksSander2015} and more recently in~\cite{NguyenBergZecchina,Cocco2017,Soding248}, and an exciting perspective has opened that such predicted contacts can also be leveraged to predict complete 3D protein structure 
\textit{in silico}~\cite{Hopf-Cell2012,Hayat2015,Baker2017,Baker2017b,Michel2017b}. Other recent applications of DCA have been to predicting RNA structure~\cite{DeLeonardis2015,DeLeonardis2017,Weinreb2017}, inter-protein contacts~\cite{Gueudre2016,Uguzzoni2017}, and synergistic effects of mutations (epistasis) not necessarily related to spatial contacts~\cite{Figliuzzi2016,Hopf2017,GenomeDCA2017}.

The actual use of DCA is comprised of two parts: (1) to run the inference procedure of choice on an MSA which consists of $N$ samples, each of dimension $L$; and (2) to keep only a subset of largest predictions for further assessment and use. We will refer to the columns of the MSA as \textit{loci}, the variables in each column as \textit{alleles}, and the rows as \textit{samples}. The MSA hence consists of $N$ samples with each sample being a list of $L$ alleles. Alternatively we will refer to $L$ as the \textit{data dimension} and $N$ as the \textit{sample size}. The total number of parameters in the inferred Ising or Potts models is proportional to $L^2$ and 
will here be denoted $\mathcal{P}$. The number of retained predictions will be denoted $\mathcal{K}$.

Exact frequentist or Bayesian-point estimate methods, \textit{i.e.} maximum likelihood (ML) or maximum a posteriori (MAP), are not computationally feasible for the data dimensions of current practical interest, and many approximate inference methods have therefore been developed~\cite{NguyenBergZecchina}. Additionally statistical identifiability demands that $\mathcal{K}$ cannot be larger than $N$; one cannot learn more features from the data than there are examples. This has indeed mostly been the case in the examples above. However, in the intermediate step $\mathcal{P}$ parameters are inferred, and in many (if not all) cases of interest $\mathcal{P} \sim L^2$ has been much larger than $N$. On top of inference being approximate it must therefore also be regularized.

The setting where DCA is applied is rather far from classical statistical inference which is mainly concerned with the limit when $N$ (number of samples) is much larger than $\mathcal{P}$ (number of parameters). Theoretical results on consistency which pertain to~\textit{e.g.} pseudo-likelihood maximization (see below) do not in themselves establish the practical superiority of this or other approaches to DCA when combined with regularization; such a conclusion instead relies on empirical testing, or on considerably more involved arguments~\cite{NIPS2016_6375,Lokhov2016,Berg2016}.

The search for methods to choose $\mathcal{K}$ given an MSA of given size and statistical characteristics was only initiated recently~\cite{Wozniak2017,Xu2017} and is yet to be developed fully. Let us note that a commonly used rule of thumb has been to retain about as many predictions as the data dimension $L$; from a statistical point of view this is inappropriate unless the MSA is square or a thin matrix, \textit{i.e.} unless $N$ is at least as large as $L$. For the protein problem this has usually been the case, but for the genome-scale problems in~\cite{GenomeDCA2017} the MSA was a fat matrix of $N$ about $10^3$ and $L$ about $10^5$. The fraction $\mathcal{K}/\mathcal{P}$ of predictions that could be retained was in this case hence less than $10^{-8}$. For most realistic genome-wide DCA problems the fraction of retained predictions would similarly be very small. In an extreme extrapolation that the genome of every living human being on earth were accurately sequenced $N$ would be $10^{10}$, while $L$, if approximately every eighth nucleotide would vary, as has been estimated for the protein-coding part of the human genome~\cite{Human-variability2016}, would be about $5\cdot 10^8$. The resulting MSA would be thin, but the number of parameters $\mathcal{P}$ of the Potts model would be very large (about $2\cdot 10^{18}$), and $\mathcal{K}/\mathcal{P}$ could again be not more than about $10^{-8}$. The scenario that only a very small fraction of predictions are retained in DCA will therefore likely remain relevant.

The issue we address in this work is the following. In practice it is cumbersome to run even approximate inference methods on the largest datasets that are of interest and if we imagine scaling up DCA to problems of the size of the human genome it will not be possible at all in the foreseeable future. However, as discussed above, inference is only used to retain a very small set of leading predictions so it is conceivable that the problem can be dimensionally reduced before inference. We will introduce a straight-forward scheme that makes this possible, and show that it works on both \textit{in silico} and real data.

The paper is organized as follows. In Section~\ref{sec:DCArev} we review the DCA approach and the pseudo-likelihood maximization (PLM) computational scheme, and in Section~\ref{sec:proposed-method-1} we formally introduce Correlation-Compressed Direct Coupling Analysis (CC-DCA) as a new inference procedure. We then test CC-DCA on \textit{in silico} data, where the models and the principles are discussed in Section~\ref{sec:test-sets}, and the results are presented in Section~\ref{sec:results}. In Section~\ref{sec:epistasis} we present an example where epistatic couplings are inferred from a collection of whole-genome sequences of the human pathogen \textit{Streptococcus pneumoniae}. We show that CC-DCA finds essentially the same leading couplings as a much more demanding DCA-based method~\cite{GenomeDCA2017} (see also \cite{PuranenDCA2017}).

On a final note the currently best-performing versions of DCA for the protein contact predictions application are hybrid schemes that rely also on other information~\cite{Jones2014,NIPS2016_6488,Wang2017,Michel2017}. Although such schemes can probably be expected to outperform ``pure DCA" also in other applications, we are in this paper only concerned with the performance of DCA and DCA-like procedures alone.

\section{A brief summary of direct coupling analysis (DCA)}
\label{sec:DCArev}

The basic assumption behind DCA is that samples are drawn from a probabilistic model of the Potts model type involving two-body interactions
\begin{equation}
  P(\bm{\sigma};\bm{J}) = \frac{1}{Z} 
  \exp\Bigl( \beta\sum_i h_i(\sigma_i) + 
  \beta\sum_{i, j} J_{i j}(\sigma_i,\sigma_j) \Bigr) \; .
  \label{eq:prob}
\end{equation}
Here $\bm{\sigma} \equiv (\sigma_1, \sigma_2, \ldots, \sigma_L)$ denotes a configuration of the system, and $\sigma_i$ is the allele (or state) of locus $i$; $\bm{h} \equiv \{h_i : i \in [1, L]\}$ and $\bm{J} \equiv \{J_{i j} : i, j \in [1, L]\}$ denotes the set of external fields and pairwise couplings; the parameter $\beta$ (inverse temperature) is introduced for later convenience and here just sets an overall scale of the energy terms. In this and the next section we will for simplicity take the $\sigma_i$'s Boolean variables (Ising spins) in the Ising gauge~\cite{WeigtWhite2009,Magnus2014} such that $h_i(\sigma_i)=h_i \sigma_i$ and $J_{i j}(\sigma_i,\sigma_j)=J_{i j}\sigma_i \sigma_j$; and we will focus on the couplings, so all the $h_i$ parameters will be zero. Furthermore all diagonal coupling elements $J_{i i}$ are set to be zero.

Given $N$ observed samples $\bm{\sigma}^{(1)}$, $\bm{\sigma}^{(2)}$, $\ldots$, $\bm{\sigma}^{(N)}$, maximum likelihood inference means to minimize the following objective function
\begin{eqnarray}
f(\bm{J})
  & \equiv & - \frac{1}{N} \sum_{n=1}^{N} \log P(\bm{\sigma}^{(n)};\bm{J}) \nonumber \\
  &=& \log Z(\bm{J}) - \beta \sum\limits_{i, j} J_{i j} \bigl\langle \sigma_i \sigma_j  \bigr\rangle \; ,
\end{eqnarray}
where $\langle \cdot \rangle$ means averaging over all the $N$ sample configurations. The optimal value of the parameter $\bm{J}$ is determined by the variational conditions
\begin{equation}
  \frac{1}{Z} \frac{\partial Z}{\partial J_{i j}}
  = \beta \bigl\langle \sigma_i \sigma_j  \bigr\rangle \; .
\end{equation}
This requires the calculation of the partition function $Z$ and its first derivatives, therefore renders the inference problem only feasible for small systems.

A part of the approach to be introduced below is pseudo-likelihood maximization (PLM)~\cite{Besag1975,NguyenBergZecchina}. This method maximizes conditional probabilities for each variable separately, and uses all the $N$ individual sample configurations in the inference process. It is statistically consistent \textit{i.e.} gives almost surely the same result as exact maximum likelihood on infinite data, but it is a weaker procedure meaning that the scatter around the true value is larger for finite data. By pseudo-likelihood the feasible size of data is extended significantly. In the case of Ising variables ($\sigma_i \in \pm 1$ for all the loci $i\in [1,L]$), the conditional probabilities for the model with energy $E(\bm{\sigma}) = -\sum J_{i j} \sigma_i \sigma_j$ is 
\begin{equation}
P\bigl( \sigma_i \ | \  \bm{\sigma}_{\backslash i} ; \bm{J} \bigr)  
= \frac{\exp(\beta \sigma_i \theta_i)}{2 \cosh(\beta \theta_i)} 
  = \frac{1}{1+\exp(-2\beta \sigma_i \theta_i)}
 \; ,
\end{equation}
where $\theta_i = \sum_j J_{i j} \sigma_j$ is the instantaneous field on locus $i$, and $\bm{\sigma}_{\backslash i} \equiv \bm{\sigma}\backslash \sigma_i$ denotes the state of all the other $(L-1)$ loci except locus $i$. The corresponding objective functions in pseudo-likelihood maximization is
\begin{equation}
f_i^{\rm{PLM}}(\bm{J})
  = \log\bigl( 1+\exp(-2\beta \sigma_i \theta_i) \bigr)  \; , \quad (i=1,\ldots, L) \; .
 \label{eq:fiPLM}
\end{equation}

We can minimize these $L$ objective functions simultaneously (``symmetric PLM''), or separately by removing the constraint $J_{i j}=J_{j i}$ (``asymmetric PLM''). Asymmetric PLM can be implemented in parallel and allows for considerable computational speed-up as the optimization problems are also smaller. However, the separate optimizations will usually give $J_{i j} \neq J_{j i}$. Following~\cite{Magnus2014} we here take the output of asymmetric PLM to be $J_{i j}^{\rm{PLM}} \equiv \frac{1}{2} (J_{i j} + J_{j i})$. We also use $l_2$ regularization as described in~\cite{Magnus2014}. Unless otherwise stated the regularization parameter $\lambda_J$ has been set to be $0.01$.

\section{Data compression before inference}
\label{sec:proposed-method-1}

Here we formally introduce Correlation-Compressed Direct Coupling Analysis (CC-DCA). The basic motivation is that although pseudo-likelihood maximization and other approximate inference methods can handle systems much larger than those for which full maximum likelihood is feasible, they still cannot be applied to very large systems. Therefore further approximations and/or simplifications are called for.

\begin{figure}
\begin{center}
\includegraphics[width=0.95\linewidth]{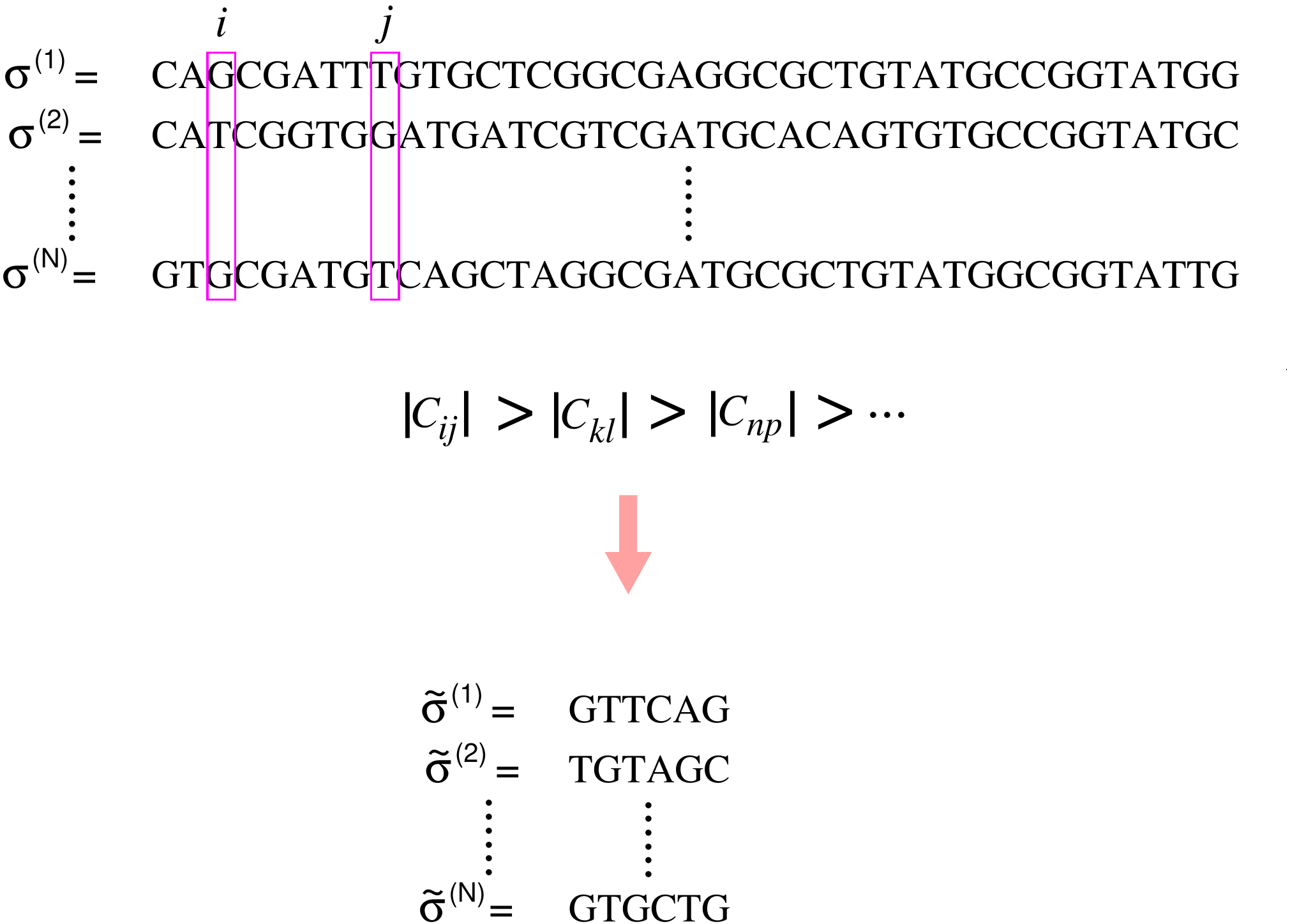}
\end{center}
\caption{\label{fig:CCidea}
Illustration of Correlation-Compressed Direct Coupling Analysis (CC-DCA) as applied to nucleotide sequence data. The input multiple sequence alignment (MSA) dataset is a $N \times L$ matrix $A$, with $N$ being the total number of sample sequences ($\bm{\sigma}^{(\alpha)}$, $\alpha=1, 2, \ldots, N$) and $L$ being the length of each sequence. Each entry is a sequence letter (one of the four nucleotides A, G, C and T). In the data analyzed in Section~\protect\ref{sec:epistasis} one of the sequence letters can also be N (anything). The covariance matrix $C_{i j}$ between any two loci $i$ and $j$ is then computed by reading the $i$-th and the $j$-th column of matrix $A$, and the column pair $(i,j)$ is given a score. In Section~\protect\ref{sec:epistasis} the score used is Mutual Information (MI). The $m$ column pairs of largest scores are selected. After considering the $\ell \leq 2 m$ loci involved in these $M$ pairs, the original MSA matrix $A$ is reduced to the new $N \times \ell$ MSA matrix $B$ for further analysis by some version of DCA.
}
\end{figure}

Our approach is to first reduce the MSA based on measured correlations, and only then apply a DCA method such as PLM. The idea is hence to take ``Direct'' in the acronym DCA both seriously and literally: two loci with a strong coupling $J_{i j}$ will in general be highly correlated, but the opposite is not true: two loci can be highly correlated without there being a strong coupling $J_{i j}$. This suggests that if the loci involved in strong correlations are retained and the other loci are eliminated, then the resulting smaller data matrix will carry enough information to determine the largest $J_{i j}$'s. The principle of CC-DCA is illustrated in Fig.~\ref{fig:CCidea}.

A list-based presentation of CC-DCA, appropriate for Ising data, is as follows:
\begin{enumerate}
\item Given an $N\times L$ MSA data matrix $A$, first compute the covariance matrix $C$ with $C_{i j} = \langle \sigma_i \sigma_j \rangle - \langle \sigma_i \rangle \langle \sigma_j \rangle$ being the correlation between the two loci $i$ and $j$.
\item Find the $m$ largest elements (either positive or negative) of the matrix $C$; and then identify the $\ell$ loci which appear in these $m$ elements. Obviously $\ell \leq 2 m$.
\item Retain these $\ell$ loci and eliminate all the others. The original MSA matrix $A$ is then reduced to a smaller $N \times \ell$ MSA matrix $B$. This correlation-compressed matrix $B$ then serves as input for DCA analysis.
\end{enumerate}

We call DCA on the correlation-compressed data CC-DCA. When the flavor of DCA is PLM we thus alternatively call the resulting algorithm Correlation-Compressed Pseudo-Likelihood Maximization (CC-PLM). This is the flavor assessed and used in the following sections.

\section{Test sets and evaluation procedures}
\label{sec:test-sets}

The study of ``inverse Ising'' and ``inverse Potts'' problems began about a decade ago stimulated by early results in neuroscience~\cite{Scheidman2006}, before it was widely appreciated that the success of DCA on practical data sets rests \textit{both} on the inference procedure \textit{and} on the choice to retain only some largest predictions. Controlled tests were done generating data from some known distribution and then checking how well the parameters could be recovered. Most of these tests were done using the root-mean-square (RMS) criterion and $\bm{J}$-matrices generated by some variant of the Sherrington-Kirkpatrick (SK) model, see \cite{Roudi2009} for an early review, and  \cite{SessakMonasson2009,AurellEkeberg2009} for two representative examples at the time and~\cite{NguyenBergZecchina} for a recent comprehensive survey.

Neither of these choices is however suitable as to how DCA methods are currently used.
The RMS criterion includes all predictions in the $J$-matrix, and not only the largest, and therefore does not reflect how well a method recovers the leading couplings. In a standard SK model on $L$ spins with Gaussian couplings the typical values of the couplings scale as  $\frac{1}{\sqrt{L}}$ and the largest values follow a Gumbel extreme value distribution with size about $\sqrt{\log L}/\sqrt{L}$. All the couplings are then of very similar values, so that this should in fact be a very challenging case, possibly much more so than realistic data. We will for completeness and back-compatibility also consider this model, but current practice in DCA additionally calls for other model classes and other evaluation criteria, as we will now discuss. 

\subsection{Random power-law test model class}
\label{sec:test-model}

As a new test model class, relatively simple to describe, we propose the random power-law model class (RPL), as follows:
\begin{itemize}
  \item The magnitudes of the elements of $\bm{J}$ are generated according to a power-law distribution, with a probability density function $\rho(x) = c \cdot x^{-\gamma}$ for $x$ in some interval $x_s\le x\le x_l$. The exponent $\gamma$ is tunable and $c$ is a normalization constant. If $\gamma > 1$ then $c= (1-\gamma)/[x_l^{1-\gamma} - x_s^{1-\gamma}]$.
  \item The signs of the elements of $\bm{J}$ can be chosen either all positive (ferromagnetic-like model), or randomized (spin-glass-like model).
  \item All elements are generated as independently and identically distributed (i.i.d.) random variables with the above characteristics.  For the Ising model ($q=2$) this just means that the coupling coefficients $J_{ij}$ are i.i.d. random variables as above while for Potts models ($q>2$) we take all the elements of the $q\times q$ coupling matrix between any two loci as i.i.d. random variables.
\end{itemize}
Obviously many similar distributions could be considered, ~\textit{e.g.} relaxing the biologically questionable assumption that the elements in a $q\times q$ coupling matrix 
are independent, but such extensions will be left for future work. The essence of the RPL class is that the coupling constants are widely distributed in size.

\subsection{SK test model class}
\label{sec:SK-test-model}

We also consider the more traditional SK spin glass model, where conventional DCA methods have been extensively tested \textit{in silico} in the past. In Section~\ref{sec:results} we apply CC-PLM on SK model data. The coupling constants $J_{i j}$ in this model will be quenched i.i.d. Gaussian random variables with mean zero and variance $\frac{1}{L}$. The coupling constants are hence narrowly distributed around zero, there are no exceptionally strong interactions. 

\subsection{Evaluation by scatter plot}
\label{sec:scatter-plot}

When testing DCA procedures on \textit{in silico} data, a most natural graphical procedure is by scatter plot. By this measure the inferred value of an interaction coefficient is given as ordinate (value on $y$-axis), plotted against the true value given as abscissa (value on $x$-axis). If the inference procedure is accurate the points will lie along the diagonal ($x=y$). If there are systematic differences between large interactions and other interactions, as there will be in the test cases described below, this will show up as deviations of the data cloud from the diagonal.

\subsection{Evaluation by true positive rate}
\label{sec:tpr}

A general evaluation procedure when using DCA on real data was introduced in~\cite{WeigtWhite2009} and has been used since in most empirical work and DCA applications. It has however not been equally used in testing on model classes, and we will therefore introduce it formally:
\begin{itemize}
\item Generate coupling coefficients/matrices $J_{i j}$ according a preferred scheme, in our case the RPL or SK as in the preceding subsections.
\item Draw $N$ independent samples from the Gibbs-Boltzmann distribution with those model parameters. In practice this has to be done with Markov-chain Monte Carlo (MCMC) and may have issues with convergence for strongly coupled systems (low temperature). In the tests below we will therefore limit ourselves to weakly coupled systems (high temperature).
\item Consider the two lists
\begin{eqnarray}
  {\cal J}^{\hbox{true}} &=&| J_{i j}^{\hbox{true},1}| \ge  |J_{i  j}^{\hbox{true},2}|\ge \cdots |J_{i j}^{\hbox{true},k}| \; , \nonumber \\
  {\cal J}^{\hbox{pred}} &=& |J_{i j}^{\hbox{pred},1}|\ge |J_{i j}^{\hbox{pred},2}|\ge \cdots |J_{i j}^{\hbox{pred},k}| \nonumber
\end{eqnarray}
of the $k$ strongest true interactions and the $k$ strongest predicted interactions and where $|\cdot |$ is a suitable norm. Compare the lists and determine how many elements $l$ they have in common. The \textit{True Positive Rate} (TPR) of the $k$ strongest couplings is then defined as
\begin{equation}
\label{eq:TPRk}
  \hbox{TPR}(k) \equiv \frac{l}{k} \; .
\end{equation}
\end{itemize}

\subsection{Evaluation by visualization}
\label{sec:evaluation-visualization}

In Section~\ref{sec:epistasis} below we consider real data where
the true couplings are unknown. In fact, it is then not known if it is 
a good approximation to assume that the data has been generated from a Potts model,
or what model class describes the data at all.
In an recent paper~\cite{GenomeDCA2017} couplings were inferred
by a modified DCA procedure and then discussed from the view-point
of plausibility and relevance in the light of how the
data had been obtained and known facts of \textit{S. pneumoniae} biology.
In this work we compare results from CC-DCA to those of~\cite{GenomeDCA2017}
by a visual procedure where couplings are displayed as arcs in a circular
plot and the darkness of an arc is proportional to coupling strength. 
The strongest inferred couplings thus stand out as isolated black arcs
on a grey background formed by many weaker inferred coupling. The circular
plots are produced by circos (\url{http://circos.ca}) \cite{Krzywinski-2009a}.
Given a list of scored couplings, evaluation by visualization proceeds as follows 
\begin{itemize}
	\item
	Partition the whole genome into non-overlapping windows of size $100$ bp. 
	\item
	Couplings connected between the same two windows are merged by a coarse-grained coupling, the score of which is simply the sum of scores before merging. The two endpoints of coarse-grained coupling are the beginning positions of the two windows.
\end{itemize}

\subsection{Evaluation of CC-DCA on \textit{in silico} data}
\label{sec:specific-problem}

We evaluate our CC-DCA method as follows. First we generate coupling coefficients/matrices according to model test classes such as RPL of Section~\ref{sec:test-model} or SK of Section~\ref{sec:SK-test-model}, and then we generate $N$ independent samples from the Gibbs-Boltzmann distribution. This yields an $N\times L$ MSA which we call data matrix $A$. We apply DCA on $A$ to get the values of all couplings, and compute a true positive rate $\hbox{TPR}^{A}(k)$. For this to be feasible $L$ cannot be very large, as discussed above.

The CC-DCA method consists in reducing $A$ to a smaller data matrix $B$ on which we run DCA. That leads to a new set of couplings obtained by CC-DCA, and to new true positive rates $\hbox{TPR}^{B}(k)$. The evaluation of the data reduction scheme then proceeds by comparing the couplings obtained from DCA and CC-DCA in a scatter-plot, and by comparing $\hbox{TPR}^{A}(k)$ to $\hbox{TPR}^{B}(k)$. Obviously, such an evaluation can only be done on relatively small systems as otherwise we could not run the DCA on the huge matrix $A$ at once. If it works, it would however support the idea to use the same procedure on very large data sets.

\section{Results on \textit{in silico} data}
\label{sec:results}

\begin{figure*}
  \begin{center}
    \subfigure[]{
      \includegraphics[width=0.23\textwidth]{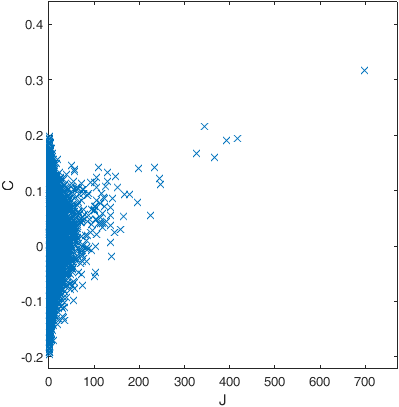}
      \label{fig:fp-a}
    }
    \subfigure[]{
      \includegraphics[width=0.23\textwidth]{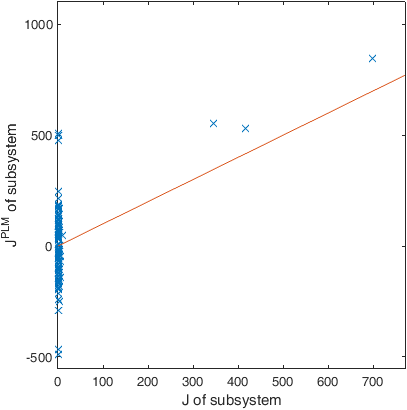}
      \label{fig:fp-b}
    }
    \subfigure[]{
      \includegraphics[width=0.23\textwidth]{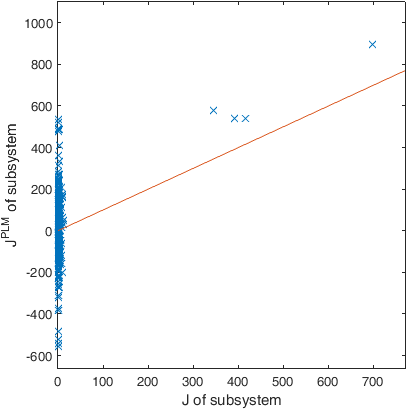}
      \label{fig:fp-c}
    }
    \subfigure[]{
      \includegraphics[width=0.23\textwidth]{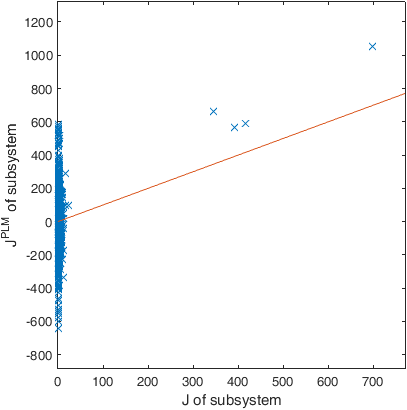}
      \label{fig:fp-d}
    } \\
    \subfigure[]{
      \includegraphics[width=0.23\textwidth]{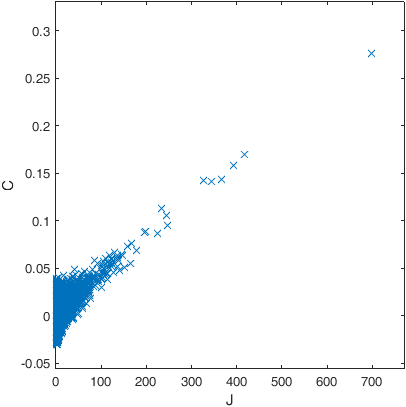}
      \label{fig:fp-e}
    }
    \subfigure[]{
      \includegraphics[width=0.23\textwidth]{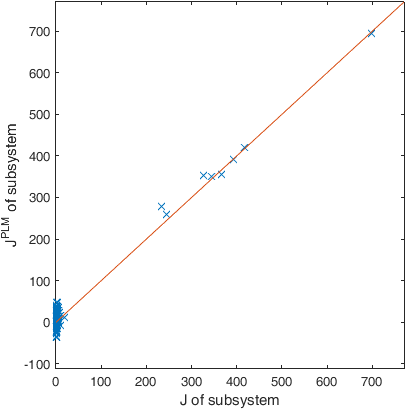}
  \label{fig:fp-f}
  }
  \subfigure[]{
  \includegraphics[width=0.23\textwidth]{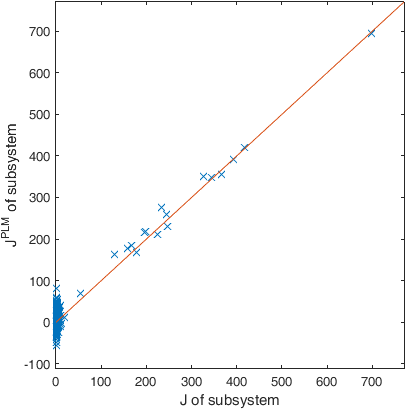}
  \label{fig:fp-g}
  }
  \subfigure[]{
    \includegraphics[width=0.23\textwidth]{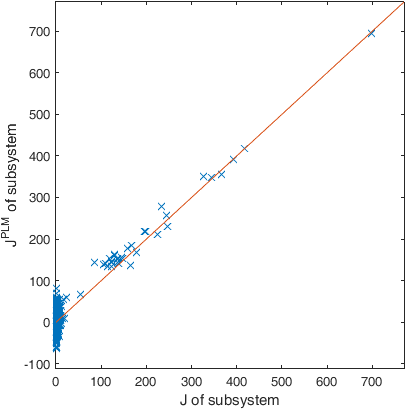}
    \label{fig:fp-h}
  }
  \end{center}
  \caption{
    \label{fig:fp-T2500-comp}
    Scatter-plot of inferred couplings vs true couplings for ferromagnetic RPL data. Number of spins is $L=1024$, number of samples (obtained at $T=2500$) is $N = 0.5 L$ for the top row [(a)--(d)] and $N=16 L$ for the bottom row [(e)--(h)]. (a) and (e): Scatter plot of covariance elements $C_{i j}$ as the inferred couplings. Other figures: scatter plot of CC-PLM results at different levels of compression as the inferred couplings. (b)--(d): CC-PLM on MSA $N=0.5L$ from subsystem of size $\ell=16$ ($m=8$ strongest covariance elements), $\ell=32$ ($m=16$), and $\ell=64$ ($m=32$), respectively. (f)--(h): same for CC-PLM on MSA $N=16L$.
  }
\end{figure*}

In this section we describe results of CC-DCA on the RPL and SK models. The data dimension $L$ is $1024$. For RPL we use power-law exponent $\gamma=3$, lower cut-off $x_s=1$ and large cut-off $x_l= \infty$. Further parameter choices are discussed together with the presentations of the results. Some additional results on the SK model with planted couplings are presented separately in Appendix~\ref{sec:SKplant}.

We schematically show results for the ferromagnetic and spin-glass couplings and for the severely under-sampled and slightly under-sampled problems and different levels of compression in the CC-DCA step. For the ferromagnetic case the signs of all the Ising terms in Eq.~(\ref{eq:prob}) are positive, while for the spin-glass case they have random signs. For the ferromagnetic model the critical temperature $T_c$ was estimated to be around $1900$ (parameter $\beta$ in (\ref{eq:prob}) about $\frac{1}{1900}$) while  for the spin-glass model the $T_c$ was estimated to be around $120$. We note again that $\beta$ is here not a physical temperature, but only sets an overall scale of the couplings. We here only report results from the high-temperature regime where $T\equiv \beta^{-1}$ is larger than $T_c$ by some margin, and so we expect that in all cases considered MCMC will converge fast enough such that the sampled configurations obey the Gibbs-Boltzmann equilibrium distribution. In the results shown here the severely under-sampled case has $N=\frac{L}{2}$ configurations, \textit{i.e.} the MSA is a fat matrix of shape $1:2$. The other case, also under-sampled, analogously has $N=16 L$; the MSA is a thin matrix of shape $16:1$. Both settings are under-sampled because $N$ is much less than the total number of model parameters $\mathcal{P}$, which is of order $L^2$.

\begin{figure*}
  \begin{center}
    \subfigure[]{
      \includegraphics[width=0.23\textwidth]{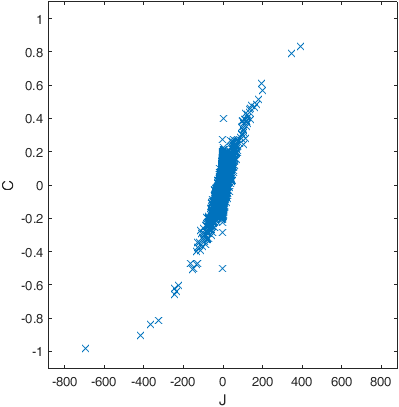}
      \label{fig:sgp-a}
    }
    \subfigure[]{
      \includegraphics[width=0.23\textwidth]{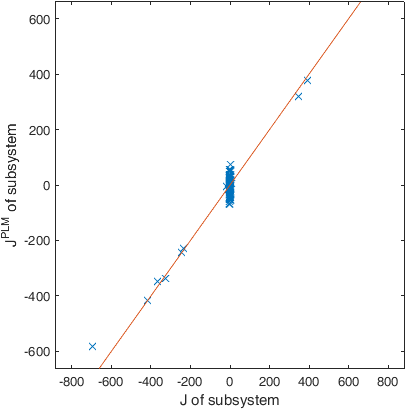}
      \label{fig:sgp-b}
    }
    \subfigure[]{
      \includegraphics[width=0.23\textwidth]{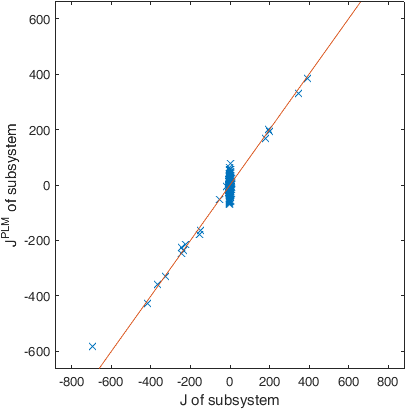}
      \label{fig:sgp-c}
    }
    \subfigure[]{
      \includegraphics[width=0.23\textwidth]{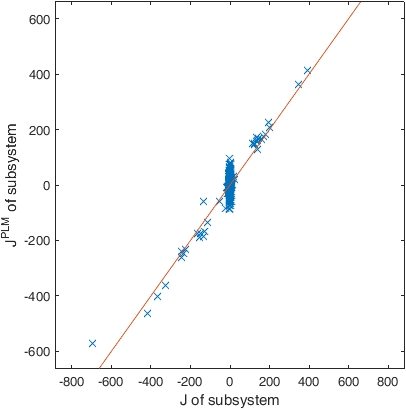}
      \label{fig:sgp-d}
    } \\
    \subfigure[]{
      \includegraphics[width=0.23\textwidth]{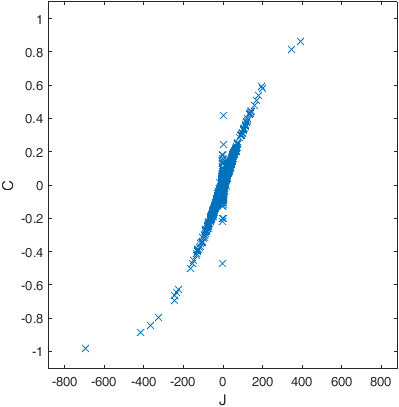}
      \label{fig:sgp-e}
    }
    \subfigure[]{
      \includegraphics[width=0.23\textwidth]{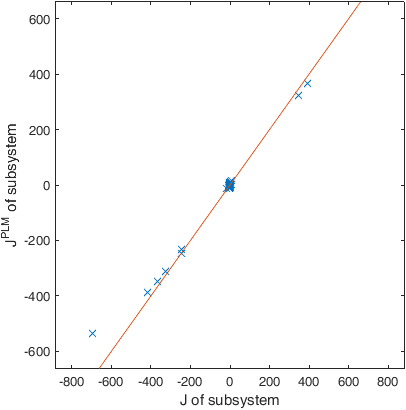}
      \label{fig:sgp-f}
    }
    \subfigure[]{
      \includegraphics[width=0.23\textwidth]{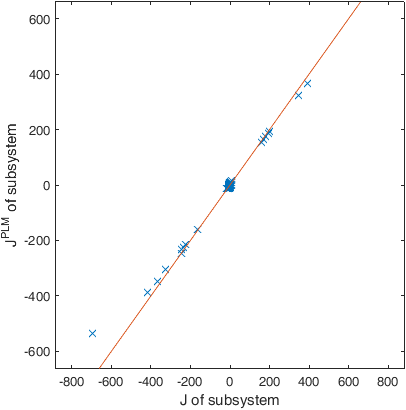}
      \label{fig:sgp-g}
    }
    \subfigure[]{
      \includegraphics[width=0.23\textwidth]{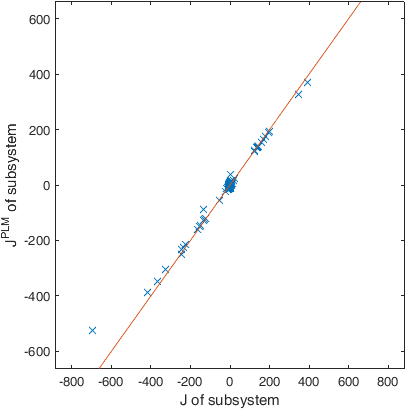}
      \label{fig:sgp-h}
    }
  \end{center}
  \caption{
    \label{fig:sgp-T300-comp}
    Same as Fig.~\ref{fig:fp-T2500-comp} but for a spin-glass RPL instance ($L=1024$ and temperature $T=300$). Number of samples is $N=0.5 L$ for the top row [(a)--(d)] and $N=16 L$ for the bottom row [(e)--(h)]. The size of the subsystem in the top row is $\ell=16$ (with $m=8$) in (b), $\ell=29$ (with $m=16$) in (c) and $\ell=57$ in (d), respectively; the size of the subsystem in the bottom row is $\ell=16$ (with $m=8$) in (f), $\ell=31$ (with $m=16$) in (g) and $\ell=56$ (with $m=32$) in (h), respectively.
  }
\end{figure*}

Fig.~\ref{fig:fp-T2500-comp} and Fig.~\ref{fig:sgp-T300-comp} show the performance of CC-DCA as scatter plots and also the performance of using bare correlations as predictors for the coupling coefficients. We see that CC-PLM has similar performance as PLM in identifying the strongest interactions in the system.  When the number of sampled configurations is relatively large (e.g., $N= 16 L$) the quantitative predictions by CC-PLM on the strongest interactions are rather accurate, even though the subsystem contains only very few loci of the original system (Fig.~\ref{fig:fp-T2500-comp}, bottom row; Fig.~\ref{fig:sgp-T300-comp}, bottom row). When the configurations are severely under-sampled (e.g., $N= 0.5 L$) there is a high danger of false positive DCA results (namely, strong interactions were predicted between some loci which actually only interact weakly); but even in this difficult case the values of the few strongest coupling constants are still predicted relatively accurately by the CC-PLM method (Fig.~\ref{fig:fp-T2500-comp}, top row; Fig.~\ref{fig:sgp-T300-comp}, top row). The couplings $J_{i j}$ in the RPL instances have quite different values and some of them are very strong (e.g., up to $J_{i j} \approx 700$). The correlations $C_{i j}$ between the  strongly interacting loci $i$ and $j$ are then naturally quite strong too. Indeed for the strongest couplings in the spin-glass case, Figs.~\ref{fig:sgp-a} and \ref{fig:sgp-e}, the
scatter plot of $C_{i j}$ vs $J_{i j}$ practically falls on a single curve,
though not on a straight line. For this reason the strongest covariance coefficients can alone serve as good indicators of the strongest direct interactions in the RPL class. 
The additional advantage of CC-PLM (and full PLM)
is that then the strengths of the strongest direct interactions can also be estimated, as 
one can see from the practically straight lines in
Figs.~\ref{fig:fp-f}-\ref{fig:fp-h} and Figs.~\ref{fig:sgp-b}-\ref{fig:sgp-d} and~\ref{fig:sgp-f}-\ref{fig:sgp-h}.

\begin{figure*}
  \begin{center}
    \subfigure[]{
      \includegraphics[width=0.4\textwidth]{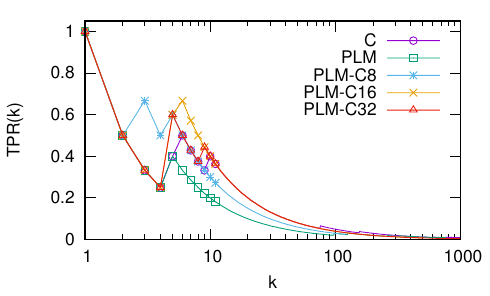}
      \label{fig:tpr-f0p5}
    }
    \subfigure[]{
      \includegraphics[width=0.4\textwidth]{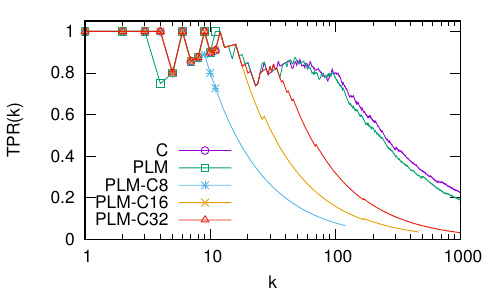}
      \label{fig:tpr-f16}
    } \\
    \subfigure[]{
      \includegraphics[width=0.4\textwidth]{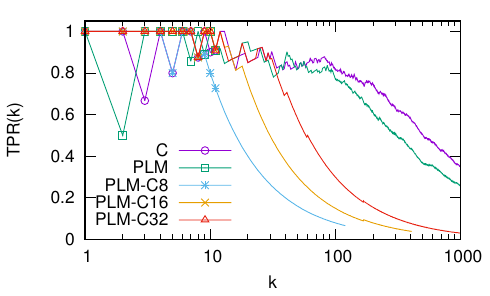}
      \label{fig:tpr-g0p5}
    }
    \subfigure[]{
      \includegraphics[width=0.4\textwidth]{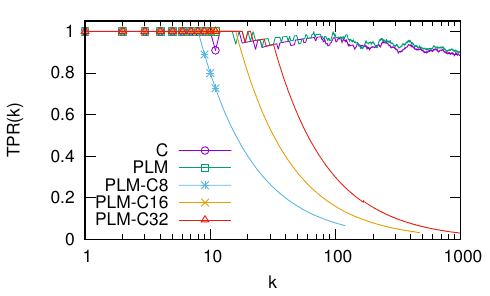}
      \label{fig:tpr-g16}
    }
  \end{center}
  \caption{
    \label{fig:tpr}
    True positive rates [Eq.~(\ref{eq:TPRk})] for random power-law model, ferromagnetic and Ising data [top row, (a) and (b)] and random power-law model, spin-glass and Ising data [bottom row, (c) and (d)]. Data dimension $L=1024$. power-law exponent $\gamma=3$. Left panels [(a) and (c)] show results with sample number $N=0.5 L$; right panels [(b) and (d)] show results with sample number $N=16 L$. Temperatures are respectively $T=2500$ for the ferromagnetic case and $T=300$ for the spin-glass case, in both cases well above the estimated critical temperature. The elements of the coupling matrix $\bm{J}$ are ranked either according to the covariance matrix $C$ (circles), or according to the PLM predictions on the whole system (squares), or according to the PLM predictions on the correlation-compressed  subsystem constructed using $m=8$ (stars), $m=16$ (crosses) and $m=32$ (triangles) strongest covariance elements.
  }
\end{figure*}

Fig.~\ref{fig:tpr} displays the same data as true positive rates.
For the severely under-sampled cases 
(Figs~\ref{fig:tpr-f0p5} and~\ref{fig:tpr-g0p5})
CC-PLM is basically able to retrieve  to the leading (largest)
couplings as well as full PLM, while for couplings
beyond the compression threshold CC-PLM falls below the other curves.
Bare correlation analysis works for these
instances almost as well as full PLM, a result that can also be
deduced from, in particular, Fig.~\ref{fig:sgp-a}.
Qualitatively the same behavior is also found for
the better sampled data (Fig.~\ref{fig:tpr-f16} and Fig.~\ref{fig:tpr-g16}).
For the better-sampled spin-glass RPL data (Fig.~\ref{fig:tpr-g16})
correlations alone are quite good predictors of the identity of the strongest coupled pairs,
a result which can also be read off from Fig.~\ref{fig:sgp-e}.
As discussed above the actual values of the couplings are less
well predicted by bare correlations, with more scatter or more 
non-linear deviations away from the diagonal in the scatter-plots
in Figs.~\ref{fig:fp-T2500-comp} and~\ref{fig:sgp-T300-comp}.

\begin{figure*}
  \begin{center}
    \subfigure[]{
      \includegraphics[width=0.3\textwidth]{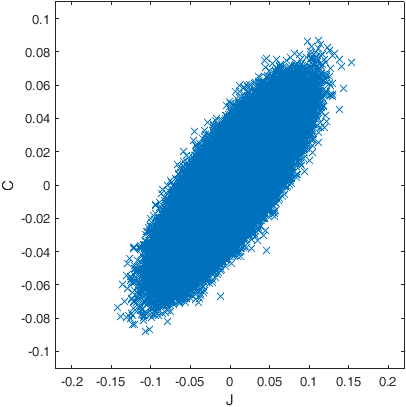}
      \label{fig:skg-a}
    }
    \subfigure[]{
      \includegraphics[width=0.3\textwidth]{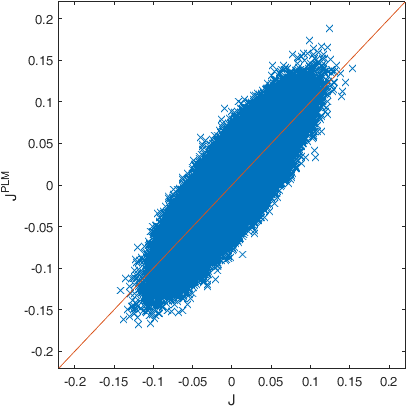}
      \label{fig:skg-b}
    }\\
    \subfigure[]{
      \includegraphics[width=0.3\textwidth]{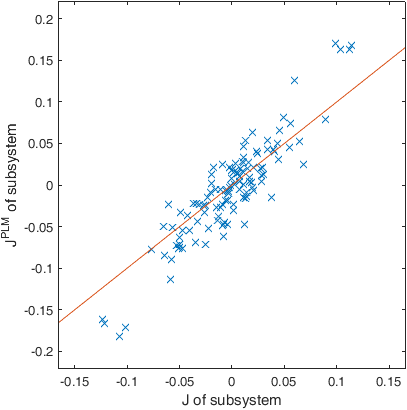}
      \label{fig:skg-c}
    }
    \subfigure[]{
      \includegraphics[width=0.3\textwidth]{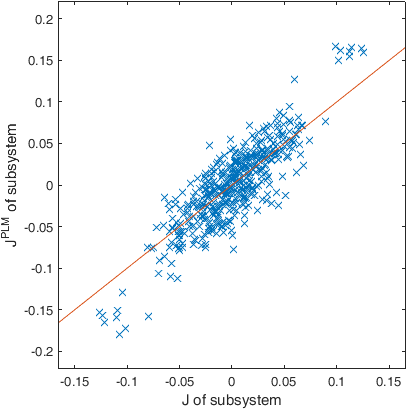}
      \label{fig:skg-d}
    }
    \subfigure[]{
      \includegraphics[width=0.3\textwidth]{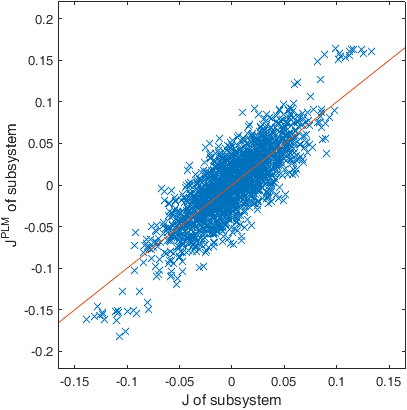}
      \label{fig:skg-e}
    }
  \end{center}
  \caption{
    \label{fig:sk-g-T2-16N-comp}
    Comparing PLM and CC-PLM on the SK spin-glass model. There are $L=1024$ spins and $p=L (L-1)/2$ coupling constants. A total number of $N= 16 L$ independent equilibrium configurations are sampled at temperature $T=2$. (a) Relation between the covariance element $C_{i j}$ and the true coupling constant $J_{i j}$. (b) Relation between the predicted coupling constant $J_{i j}^{PLM}$ and the true value $J_{i j}$ for the whole system. (c)--(e): Relation between the predicted coupling constant and the true coupling constant for the subsystem of size $\ell=16$ (obtained by considering the $m=8$ strongest covariance elements), $\ell=31$ (for $m=16$) and $\ell=62$ (for $m=32$), respectively.
   }
\end{figure*}

We then turn to applying CC-PLM to the SK spin-glass model. As Fig.~\ref{fig:skg-a} suggests, the covariance $C_{i j}$ scales roughly linearly with the coupling constant $J_{i j}$ in the high-temperature region, but there is a high degree of dispersion due to under-sampling of equilibrium configurations (here we use $N=16 L$). If the sampled configurations are used by PLM to infer the coupling constants, the qualitative agreement with the true values is better but not perfect [Fig.~\ref{fig:skg-b}]. Results in this direction were obtained already in the early DCA literature \textit{cf.} \cite{SessakMonasson2009,AurellEkeberg2009} and have recently been developed further~\cite{NguyenBergZecchina,Berg2016}. We here apply CC-PLM on the subsystem corresponding to the $m$ strongest covariance elements. The inference results for the subsystem of size $\ell=16$ (for $m=8$), $\ell=31$ (for $m=16$) and $\ell=62$ (for $m=32$) are shown in Fig.~\ref{fig:skg-c}, \ref{fig:skg-d} and \ref{fig:skg-e}, respectively. The predicted values of the coupling constants in these subsystems are in good match with the true values. The CC-PLM method therefore is capable of identifying the strongest interactions also in these systems, but 
the inferred values of the coupling coefficients are less accurate than in the RPL class.

\section{Retrieval of epistatic couplings from whole-genome bacterial data by CC-DCA}
\label{sec:epistasis}

In this Section we discuss retrieving couplings on the genome scale from real data.
The general biological term for combinatorial effects in fitness is 
epistasis~\cite{epistasis}. 
All the settings where DCA has been applied can be considered special cases of epistasis, generated by the physical interactions of residues in a protein, or by any other mechanism. 
As in the DCA literature overall we here assume that inferred Ising/Potts parameters directly measure epistasis.
The phenomenon of correlated variations between loci in data is called Linkage Disequilibrium (LD)~\cite{LD}. LD can be due both to epistasis and to shared ancestry
of loci at close enough genomic positions.
In the following we will separate long-range couplings where shared ancestry is
unlikely from short-range couplings where LD could be caused by both epistasis and shared ancestry.

In recent years datasets have been obtained on whole genomes of samples from entire bacterial populations.
Characteristic sizes of these datasets are $L$ not larger than a few millions (size of a bacterial genome) and $N$ not larger than a few thousand (largest current samples). In practice genomes in naturally occurring organisms only vary on a subset of all positions, so that the number of varying loci $L$ is not larger than on the order of one or a few hundred thousand. Still, the number of Potts model parameters to describe a distribution over $100,000$ loci would be on the order of $10^{10}$ and to learn such models directly from data is very challenging.

In two recent contributions PLM was used to analyze epistasis in the human pathogen~\textit{S. pneumoniae} (the pneumococcus). In the first approach~\cite{GenomeDCA2017} the pneumococcal genome was split into about $1500$ chunks, one locus was randomly selected from each chunk, and PLM was run on this (much reduced) set, and then run again on a new random selection, and so on. A putative interaction was scored by how many times it appeared in the lists from each sampling, which required many samplings, in practice several tens of thousands. In the second approach~\cite{PuranenDCA2017} an optimized version of PLM was run on all the loci at once, and the inferred Potts parameters used as in standard DCA. Both methods yield very similar results, but both also led to substantial computation time. We will here see how well CC-DCA manages on this challenging real-world dataset, assuming that the results from~\cite{GenomeDCA2017} can be taken as "ground truth". Evaluation will be by visual comparison as described in Section~\ref{sec:evaluation-visualization}.

\subsection{Preparation of data}
The data contains the genome alignment for $3156$ isolates of \textit{S. pneumoniae} downloaded from the data repository~\cite{GenomeDCA2017-data}. The format of this MSA are letters in the alphabet A, C, G, T and N (with N meaning complete uncertainty on the letter), each of which has length $2,221,315$. Thus the data is severely under-sampled. A Potts model fitted to this data (with gauge chosen) would have $\frac{L(L-1)}{2} (q-1)^2$ parameters for pairwise interaction and $L(q-1)$ parameters for local fields, where $q = 5$ and $L = 2,221,315$, in total about $32\cdot 10^{12}$ real parameters. 

Following~\cite{GenomeDCA2017} we first filter to remove loci that lack information and loci that are not bi-allelic. For each locus, ignoring N, we denote the most common letter (among A, C, G, T) as \textit{major} and the second most common letter as \textit{minor}. Our filtering criteria are:
\begin{enumerate}
	\item
	\textbf{Remove multi-allelic loci}. A locus is considered as multi-allelic when the counter of the third most common letter (among A, C, G, T) is not zero.
	\item
	\textbf{Remove ``frozen'' loci.} A locus is considered as frozen when its minor allele frequency (MAF) is less than $0.01$. For bi-allelic loci, the MAF is computed by
	\begin{equation}
	\text{MAF} = \frac{\text{minor}}{\text{major} + \text{minor}} \; .
	\end{equation}
	\item
	\textbf{Remove loci which have high uncertainty.} A locus is considered as highly uncertain when its frequency of the letter N is larger than $500/3156 \approx 0.158$.
\end{enumerate}
%
Among the $2,221,315$ loci, $2,177,096$ loci are bi-allelic and out of which $113,237$ loci have MAF at least $0.01$. $31,731$ of these remaining loci have high uncertainty; after removing these additional uncertain loci, $81,506$ loci survive. 
After this filtering we reduce the number of states $q$ from $5$ to $3$. The states are $N$ ($s=1$), \emph{major} ($s=2$) and \emph{minor} ($s=3$). In the context of statistical physics, the resulting MSA dataset is a collection of $3156$ configurations for a $q=3$ Potts model with $81,506$ nodes; by construction \emph{major} is the most common symbol at all loci, and we therefore (trivially) expect to find everywhere an inferred external field favouring state $2$.
Following standard procedures in DCA, also used in~\cite{GenomeDCA2017}, we apply the re-weighting procedure described in~\cite{WeigtWhite2009} and~\cite{Magnus2013,Magnus2014}. After re-weighting with threshold $x=1$ (namely, if $k\geq 2$ rows of the $3156\times 81506$ MSA matrix are identical, only one of them is kept while the other $k\!-\!1$ rows are deleted), the number of configurations went down from $3156$ to $3145$ \textit{i.e.} only a very minor change.

\subsection{Results}

\begin{figure}
  \centering
  \includegraphics[width=0.95\linewidth]{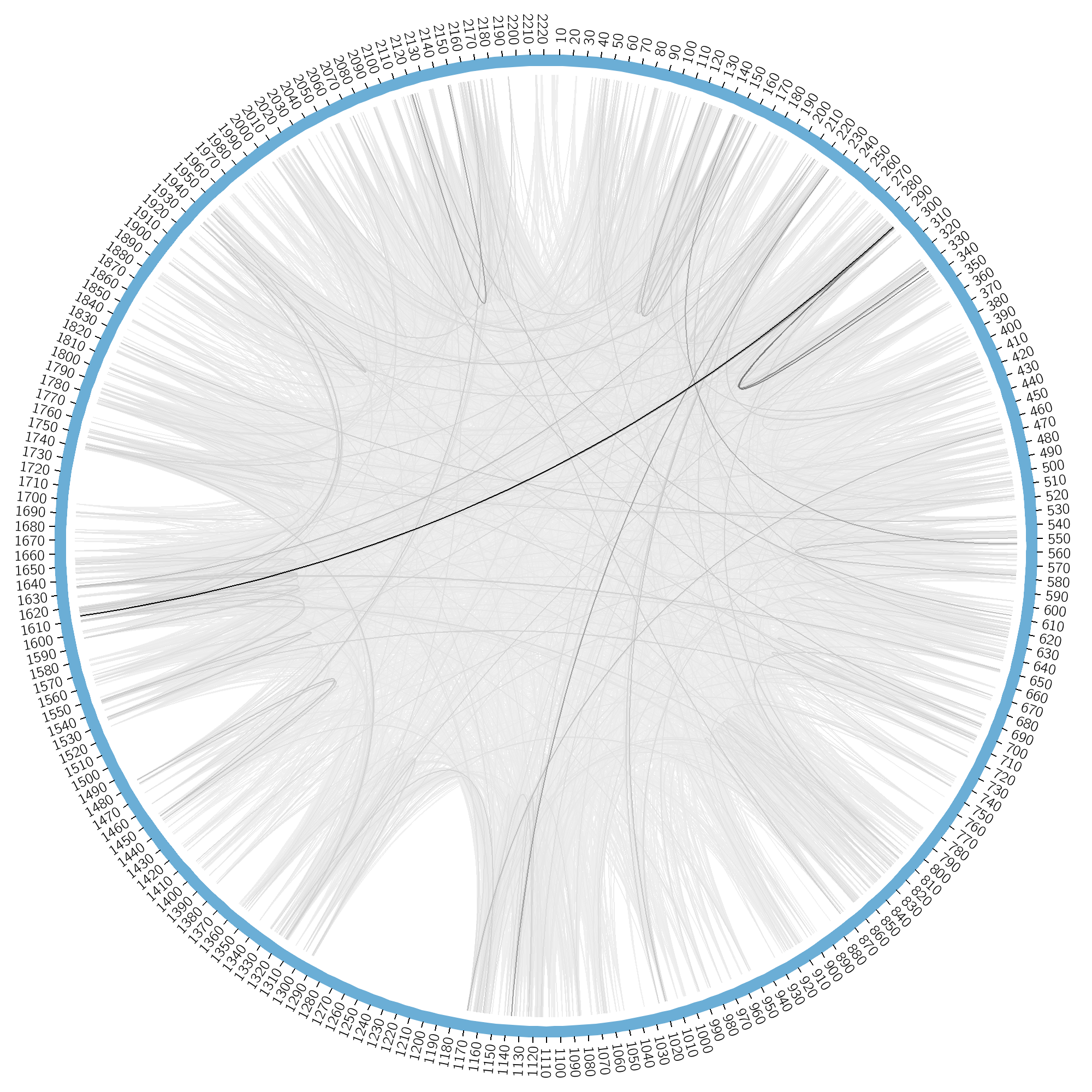}
  \caption{
    \label{fig:ccplm-30000}
    The $15717$ long-range couplings among the $10^5$ strongest couplings identified by CC-PLM with $30,000$ largest correlations. The number of loci that appear as endpoints of these links is $\ell= 9300$. Darkness of edges represent the strength of couplings. Positions along outer rim are genomic coordinates in units of $1000$ bp (described in text). Short-range couplings, the distance of which is smaller than $10,000$ bp, are not shown in this figure, but are indicated in Figs.~\protect\ref{fig:ccplm-30000b} and~\protect\ref{fig:ccplm-10000}. Coarse-graining was used in visualisation (described in text). Interactions between genes pbp2b (position $1615$), pbp2x (position $290$) and pbp1a (position $330$) are clearly visible.    
  }
\end{figure}

\begin{figure}
  \centering
  \includegraphics[width=0.95\linewidth]{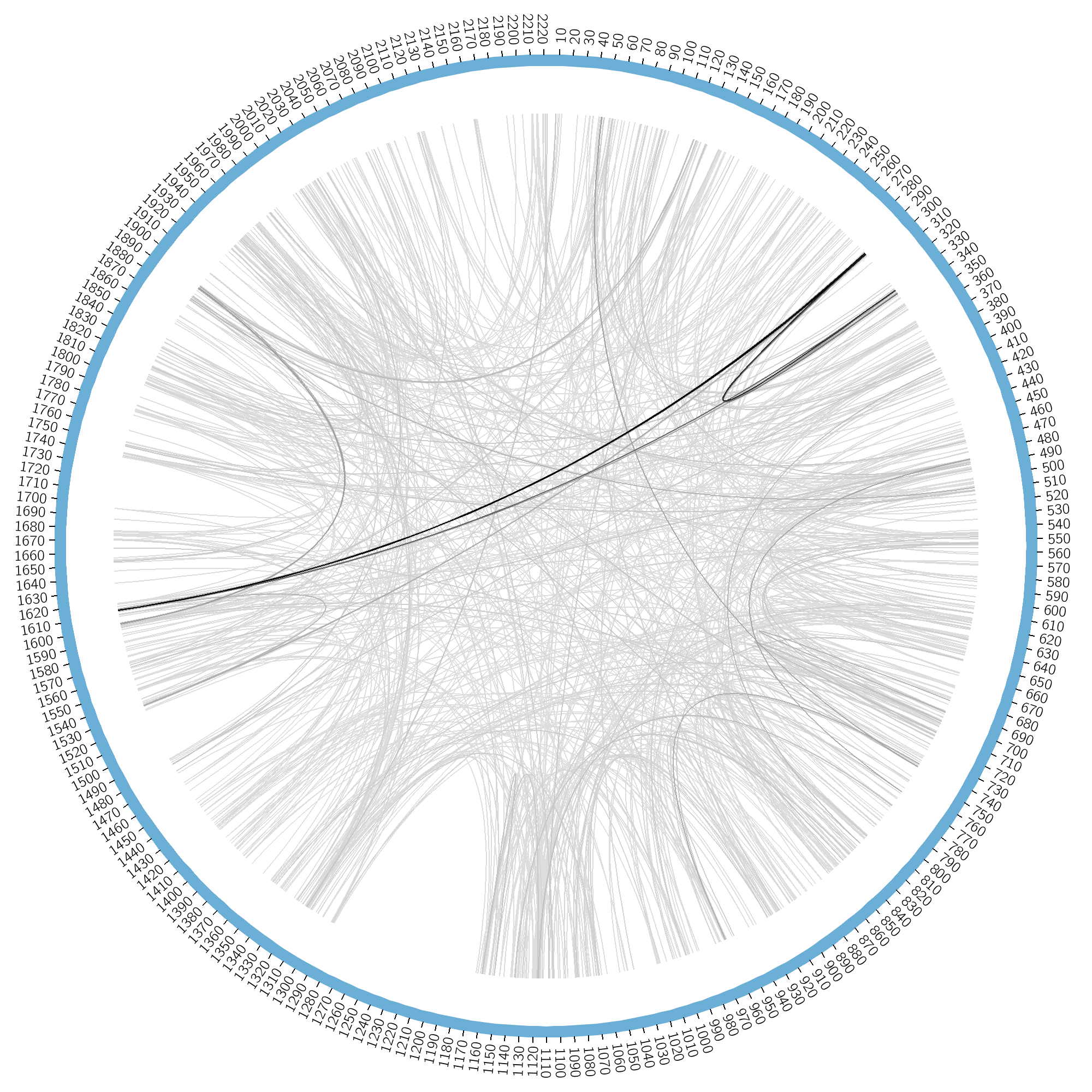}
  \caption{
    \label{fig:genomeDCA}
    The $5199$ strong couplings identified in~\protect\cite{GenomeDCA2017} depicted with the same visualisation procedure as used in Fig.~\protect\ref{fig:ccplm-30000}. Only long-range couplings are displayed. Additionally to the links between genes pbp2x, pbp2b and pbp1a one can here also identify the link to dyr (position $1530$), and the triad of interactions involving divIVA (position $1600$), psp (position $120$) and a site upstream of gene ply (position $1890$).
}
\end{figure}
To quantify correlations between two loci by a scalar we use as in~\cite{WeigtWhite2009} mutual information (MI). The first step in CC-DCA is hence to find for each pair of loci $i$ and $j$ a real number which is the absolute value of MI between these two columns in the MSA, then to order the pairs by this number in descending order, and then to identify the set of loci which are members in a list of $m$ top-ranking pairs. On this subset (MSA $B$) of loci we then run DCA.
We use the asymmetric version of plmDCA~\cite{Magnus2014} with hyper-parameters $\lambda_h = 0.1$ and $\lambda_J = 0.05$. The inferred couplings between loci $i$ and $j$
are scored by a modified Frobenius norm where the state N is not counted \textit{i.e.}
\begin{equation}
    \mathcal{S}_{i j} = \sqrt{\sum\limits_{s_i =2}^3 \sum\limits_{s_j = 2}^{3} J^2_{i j}(s_i, s_j)} \; ,
\end{equation}
where $s_i$ and $s_j$ are the states of the two loci $i$ and $j$ and the coupling matrix $J_{ij}(s_i, s_j)$ is in the Ising gauge~\cite{WeigtWhite2009, Magnus2013}. This procedure is analogous to the plmDCA20 method described in~\cite{Feinauer2014}, where only residues (not gap states) were included in the scoring, and which was there shown to improve the accuracy of contact prediction in a large test set of protein families.

\begin{figure}
  \centering
  \includegraphics[width=0.95\linewidth]{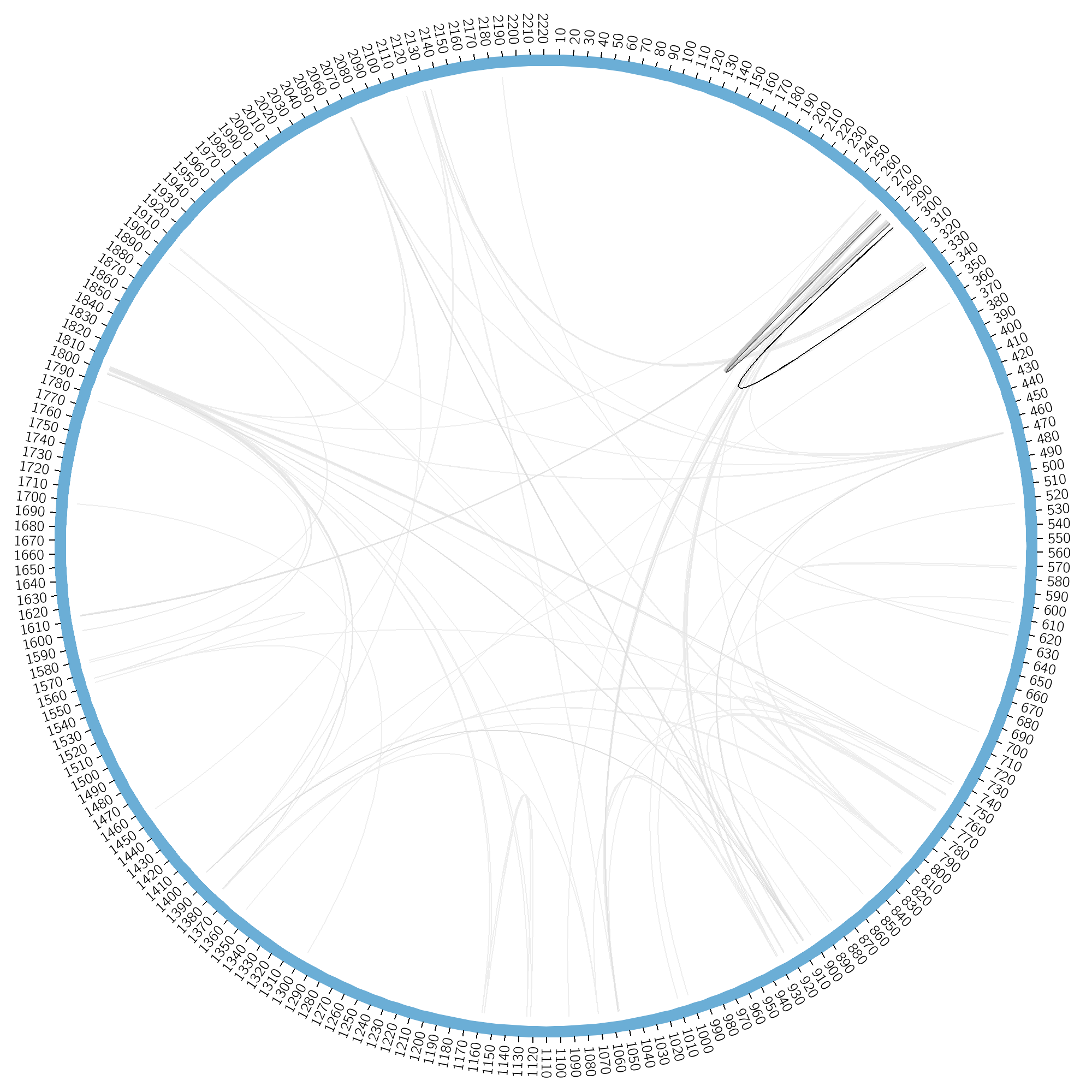}
  \caption{
  \label{fig:ccMI-30000}
  The $907$ long-range correlations among the $100,000$ strongest correlations from which the loci used to generate data in Fig.~\protect\ref{fig:ccplm-30000} were identified. Similarly to Fig.~\protect\ref{fig:ccplm-30000} one can identify the comparatively short-range interaction between pbp2x (position $290$) and pbp1a (position $330$) but the interactions to pbp2b (position $1615$) or less evident; additionally one observes in this figure a strong shorter-range inferred interaction between positions $290$ and $280$, a presumably spurious link absent in Fig.~\protect\ref{fig:genomeDCA}.
  }
\end{figure}

The results obtained by CC-DCA from $9300$ loci involved in the $30,000$
strongest correlations
are shown in Fig.~\ref{fig:ccplm-30000}, and in Figs.~\ref{fig:ccplm-30000b} and~\ref{fig:ccplm-10000}
in Appendix~\ref{sec:ccplm-10000}. 
The processing steps in the visualization are described above 
in Section~\ref{sec:evaluation-visualization}.
The numbering around the rim in Fig.~\ref{fig:ccplm-30000}
is genomic position in units of $1000$ bp. The plots hence go along the whole~\textit{S. pneumoniae} genome in our data, from genomic coordinate $1$ to $2,221,315$.
We evaluate by comparing to a visualization of the $5199$ strongest couplings identified in~\cite{GenomeDCA2017}, shown in Fig~\ref{fig:genomeDCA},
and also by a visualization of the same $30,000$
strongest correlations as such (Fig.~\ref{fig:ccMI-30000}, commented on below).
The interactions between genes pbp2b (position 1615) and pbp2x (position 290) as well as between pbp2x and pbp1a (position 330) are immediately visible 
in Figs.~\ref{fig:ccplm-30000} and~\ref{fig:genomeDCA}. It is also possible to identify 
other links discussed in~\cite{GenomeDCA2017} (see caption to Fig.~\ref{fig:genomeDCA})
as well as a characteristic absence of couplings involving loci at positions $1170-1290$. 

Turning now to correlations, 
among the $30,000$ strongest which were used in
obtaining Fig.~\ref{fig:ccplm-30000} (and Fig.~\ref{fig:ccplm-30000b}), only $907$ are long-range. These are displayed in Fig.~\ref{fig:ccMI-30000}. 
As there are now much fewer links ($30,000$, as opposed to the much larger number of
couplings inferred from the reduced MSA,
the background density (the overall grey shade of the figures) are different.
More to the point we see that Fig.~\ref{fig:ccMI-30000} 
emphasizes a relatively short-range correlations between positions $270$ and $280$, while
Fig.~\ref{fig:ccplm-30000} emphasizes long-range interaction between positions $290$ and $1610$ (both show the link between $290$ and $330$).
The link between $270$ and $280$ is not found in Fig.~\ref{fig:genomeDCA},
and also otherwise Figs.~\ref{fig:ccplm-30000} and~\ref{fig:genomeDCA}
are clearly the most similar to one another.

In conclusion, the agreement between the results obtained by 
CC-DCA and the DCA-derived method in~\cite{GenomeDCA2017} 
should be deemed fair, especially given that CC-DCA here represents a very significant simplification of the computational task. 
Although some results from~\cite{GenomeDCA2017} can also be identified
directly from correlations (Fig.~\ref{fig:ccMI-30000}), overall the agreement between CC-DCA and DCA is better.

\section{Discussion}
\label{sec:discussion}

We have in this work introduced Correlation-Compressed Direct Coupling Analysis (CC-DCA) as a convenient method to detect the strongest direct interactions from datasets (MSAs) so large that direct application of DCA is cumbersome or not feasible. We have validated this method on synthetic data sampled from the random power-law (RPL) model and standard Sherrington-Kirkpatrick (SK) model, as well as (in Appendix~\ref{sec:SKplant}) SK with some additional planted large couplings. Results are good to very good for all cases tested.

We have also shown that CC-DCA allows to recover, at very low computational overhead, results on whole-genome bacterial population-wide sequence data. Methods in the PLM family are quite challenging and resource-demanding to run on the full datasets under consideration, hence we have here only compared CC-PLM with the published results  in~\cite{GenomeDCA2017}, and not with results that could be obtained by the optimized version of PLM recently presented in~\cite{PuranenDCA2017}; we leave that to future work.

We have in this work not given detailed performance measures since the components of CC-DCA are either standard (computation of covariance matrices) or have been amply documented in the earlier literature (using PLM on a reduced data set of standard size). 
For the data sizes tested in the present paper the main computational bottleneck of CC-DCA is to compute the covariance matrix based on the empirical data. For the whole-genome MSA dataset in Section~\ref{sec:epistasis} the total time used to compute all the correlations by a single-threaded \texttt{C++} code was $15.8$ hours on a processor with frequency $2.2$ GHz and runtime memory is less than $8$ GB.
In practical applications this task can be further simplified by maintaining a running list of the $m$ strongest covariance elements $C_{i j}$ and discarding all the other elements. Further simplifications and improvements will be a future task.

In summary we have demonstrated a new means of application of DCA-like methods
to very large datasets of biological interest by using intelligent pre-processing
to reduce computational costs by a large factor.

\section*{Acknowledgement}

This research was supported by the Chinese Academy of Sciences CAS Presidents International Fellowship Initiative (PIFI) grant No. 2016VMA002 (E.A.), by the Academy of Finland through its Centre of Excellence COIN (E.A., grant no. 251170), and by the National Science Foundation of China (grant numbers 11647601 and 11421063). We thank Pan Zhang for inspiring discussion and Marcin Skwark for discussions and valuable comments on the manuscript. The numerical computation was carried out partly at the HPC cluster of ITP-CAS.

\appendix
\section{CC-DCA on SK model with planted couplings}
\label{sec:SKplant}

In this Section we consider a simple modified SK model with planted couplings in the form of two chains of strongly interacting loci. The test will then be how well CC-DCA can recover the planted loci. The system is constructed as follows: 
\begin{enumerate}
  \item Generate a graph for the SK model.  Number of spins is $L=1024$; each $J_{i j}$ is an i.i.d. Gaussian distributed random real value with mean zero and variance $L^{-1}$.
  \item Add one ferromagnetic chain of length $10$ by modifying nine coupling constants as: $J_{ij} \leftarrow J_{ij}+10, \quad {\rm{for}}\  (i,j) \in \{(1,2), (2,3), \cdots, (9,10)\}$.
  \item Add one anti-ferromagnetic chain of length $10$ by modifying nine coupling constants: $J_{ij} \leftarrow J_{ij}-10, \quad {\rm{for}}\ (i, j) \in \{(11,12), (12,13), \cdots, (19,20)\}$.
\end{enumerate}

\begin{figure*}
  \begin{center}
    \subfigure[]{
      \includegraphics[width=0.23\textwidth]{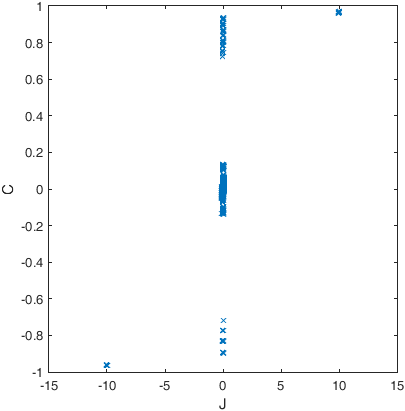}
      \label{fig:sk-pa}
    }
    \subfigure[]{
      \includegraphics[width=0.23\textwidth]{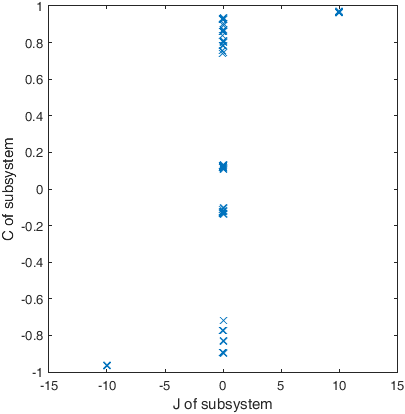}
      \label{fig:sk-pb}
    }
    \subfigure[]{
      \includegraphics[width=0.23\textwidth]{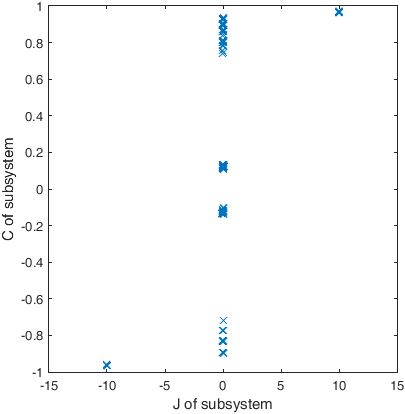}
      \label{fig:sk-pc}
    }
    \subfigure[]{
      \includegraphics[width=0.23\textwidth]{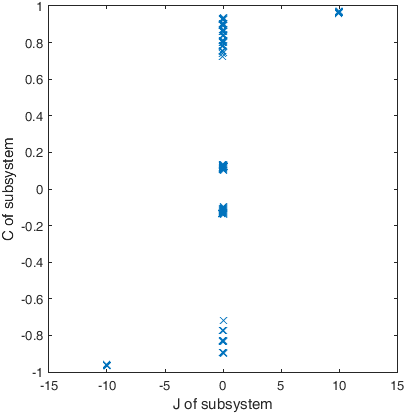}
      \label{fig:sk-pd}
    }\\
    \subfigure[]{
      \includegraphics[width=0.23\textwidth]{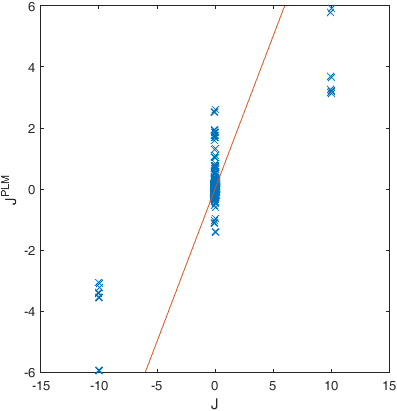}
      \label{fig:sk-pe}
    }
    \subfigure[]{
      \includegraphics[width=0.23\textwidth]{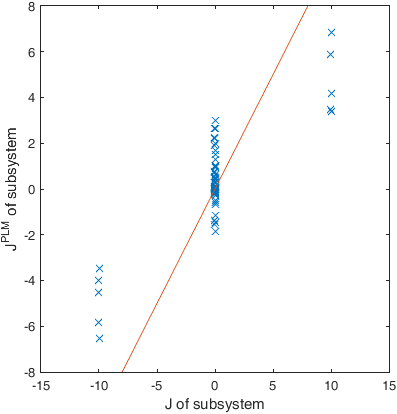}
      \label{fig:sk-pf}
    }
    \subfigure[]{
      \includegraphics[width=0.23\textwidth]{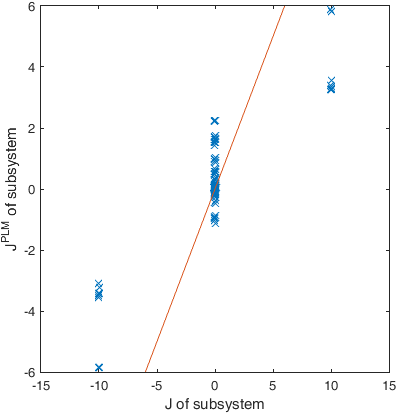}
      \label{fig:sk-pg}
    }
    \subfigure[]{
      \includegraphics[width=0.23\textwidth]{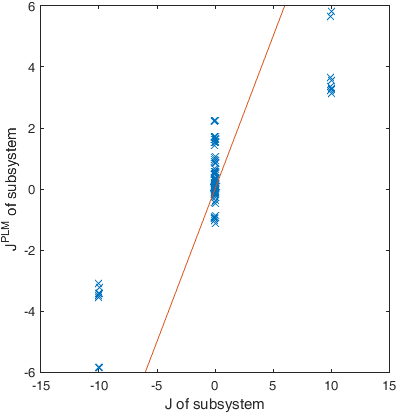}
      \label{fig:sk-ph}
    }
    \\
    \subfigure[]{
      \includegraphics[width=0.23\textwidth]{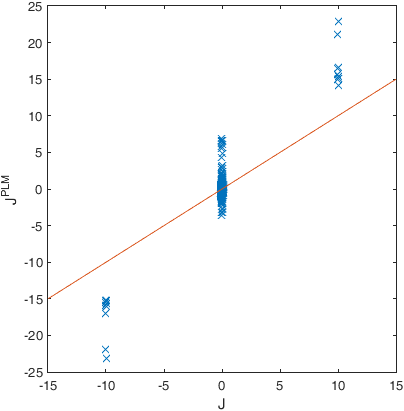}
      \label{fig:sk-pi}
    }
    \subfigure[]{
      \includegraphics[width=0.23\textwidth]{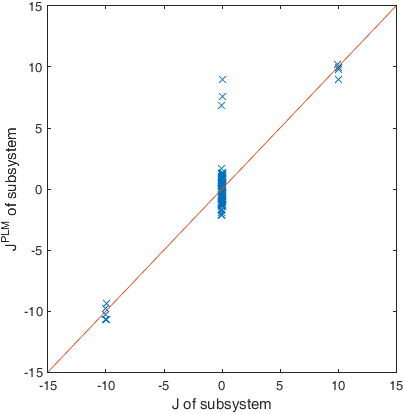}
        \label{fig:sk-pj}
    }
    \subfigure[]{
      \includegraphics[width=0.23\textwidth]{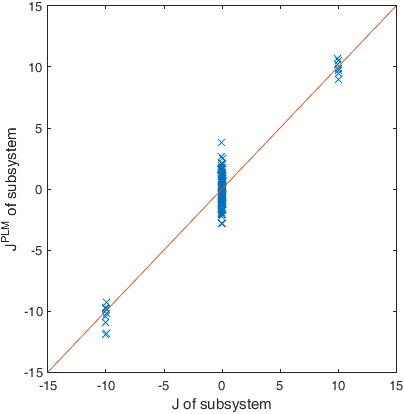}
      \label{fig:sk-pk}
    }
    \subfigure[]{
      \includegraphics[width=0.23\textwidth]{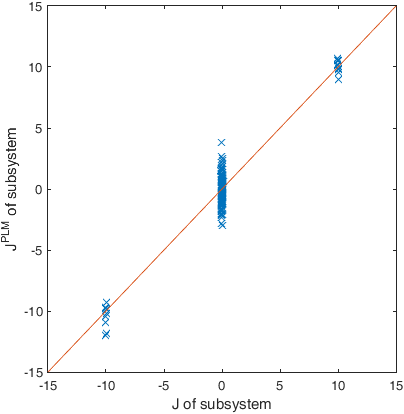}
      \label{fig:sk-pl}
    }
  \end{center}
  \caption{
    \label{fig:sk-p-T5-16N-comp}
    DCA (the first column) and CC-DCA (the 2nd to 4th columns) analysis on a chain-SK model containing $L=1024$ spins and $p = L (L-1)/2$ couplings $J_{i j}$. A total number of $N = 16 L$ equilibrium spin configurations are sampled from this model at temperature $T=5.0$. First row [(a)--(d)]: Relation between coupling $J_{i j}$ and covariance element $C_{i j}$. Second row [(e)--(h)] and Third row [(i)--(l)]: Relation between the predicted coupling $J_{i j}^{PLM}$ and the true coupling $J_{i j}$; results in the second row are obtained with regularization parameter $\lambda=10^{-2}$, while results in the third row are obtained with $\lambda=10^{-5}$.  The first column [(a), (e), (i)] corresponds to the whole system. The second column [(b), (f), (j)], the third column [(c), (g), (k)], and the fourth column [(d), (h), (l)] correspond to the subsystem with $\ell=15$ (according to the $m=8$ strongest covariance elements), $\ell=19$ (for $m=16$) and $\ell=20$ (for $m=32$) spins, respectively.
  }
\end{figure*}

Note that the coupling constants in each of the two chains are strongly correlated. The critical temperature of the conventional SK model is $T=1$. When the temperature is much higher than this value, most of the spins in the system are only weakly coupled, except those in the two chains. The spins in the two chains are strongly correlated even if they are not directly coupled with each other [Fig.~\ref{fig:sk-pa}]. We can perform DCA analysis on the $N$ sampled equilibrium configurations through PLM. This method assigns a value to each of the $\mathcal{P}= L (L-1)/2$ coupling constants. As demonstrated in Fig.~\ref{fig:sk-pe} and \ref{fig:sk-pi} the performance of this method is relatively good even when the number of sampled configurations $N$ is much smaller than the total number of parameters $\mathcal{P}$.

In the case of under-sampling ($N \ll \mathcal{P}$) the objective is not so much to infer all the coupling constants but to identify the most significant interactions. For this latter task we can construct a subsystem by retaining only the spins involved in the strongest correlations. As demonstrated in Fig.~\ref{fig:sk-p-T5-16N-comp} (2nd, 3rd, and 4th column) the CC-PLM works fine for this problem instance. It is able to distinguish the true interactions even if the subsystem only contains in the range from $15$ to $20$ spins. 

\section{Additional results on the whole-genome dataset}
\label{sec:ccplm-10000}

\begin{figure}
  \centering
  \includegraphics[width=0.95\linewidth]{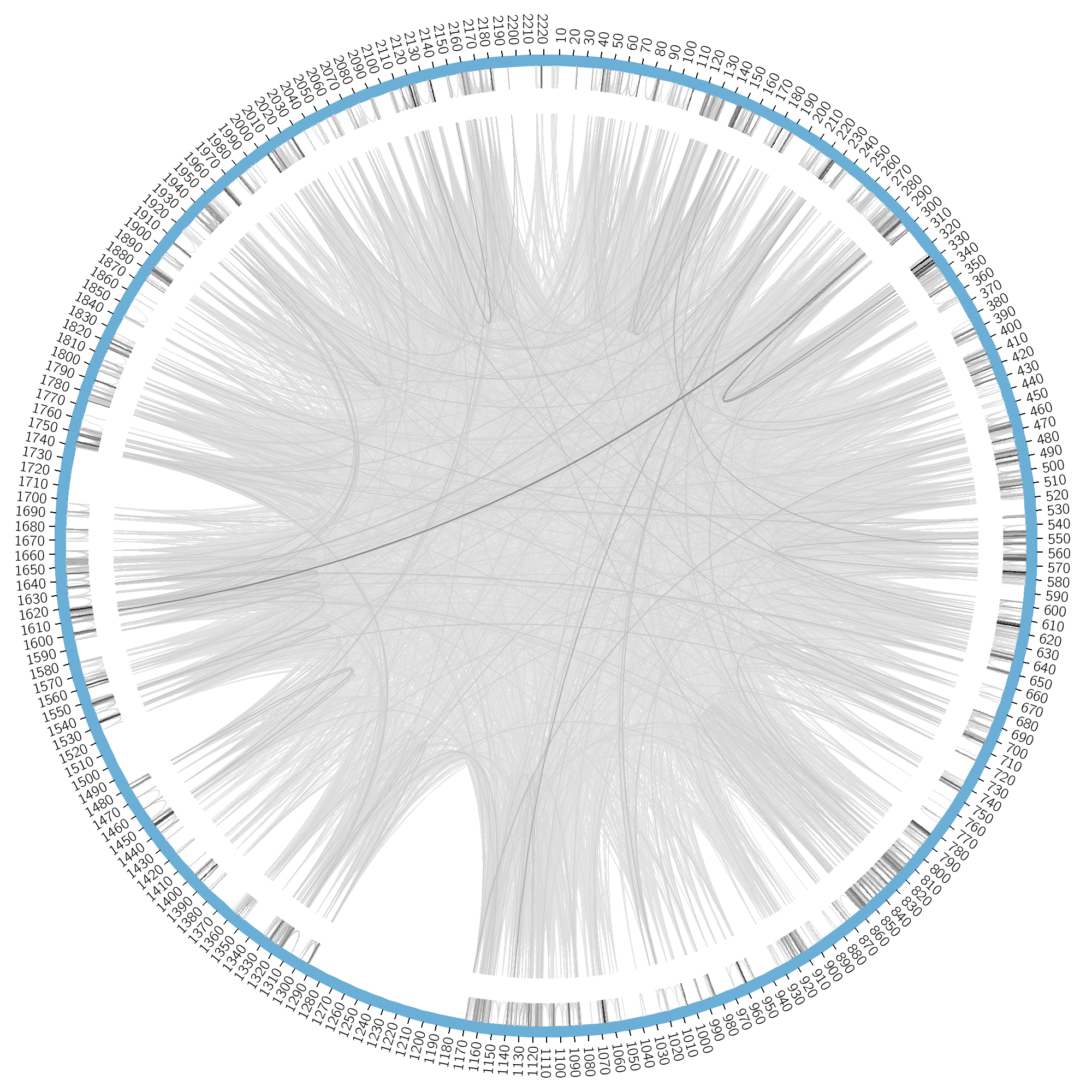}
  \caption{
    \label{fig:ccplm-30000b}
    This figure is the same as Fig.~\ref{fig:ccplm-30000} but with all the $10^5$ strongest couplings identified by CC-PLM with $30,000$ largest correlations. The number of loci involved is $9300$. Darkness of edges represent the strength of couplings. Short-range couplings, the distance of which is smaller than $10,000$ bp, are depicted around the outer rim; long-range couplings are depicted as arcs. Many short-range interactions appear in or around pbp1a (position 330) and on other positions around the genome.
    }
\end{figure}

\begin{figure}
  \centering
  \includegraphics[width=0.95\linewidth]{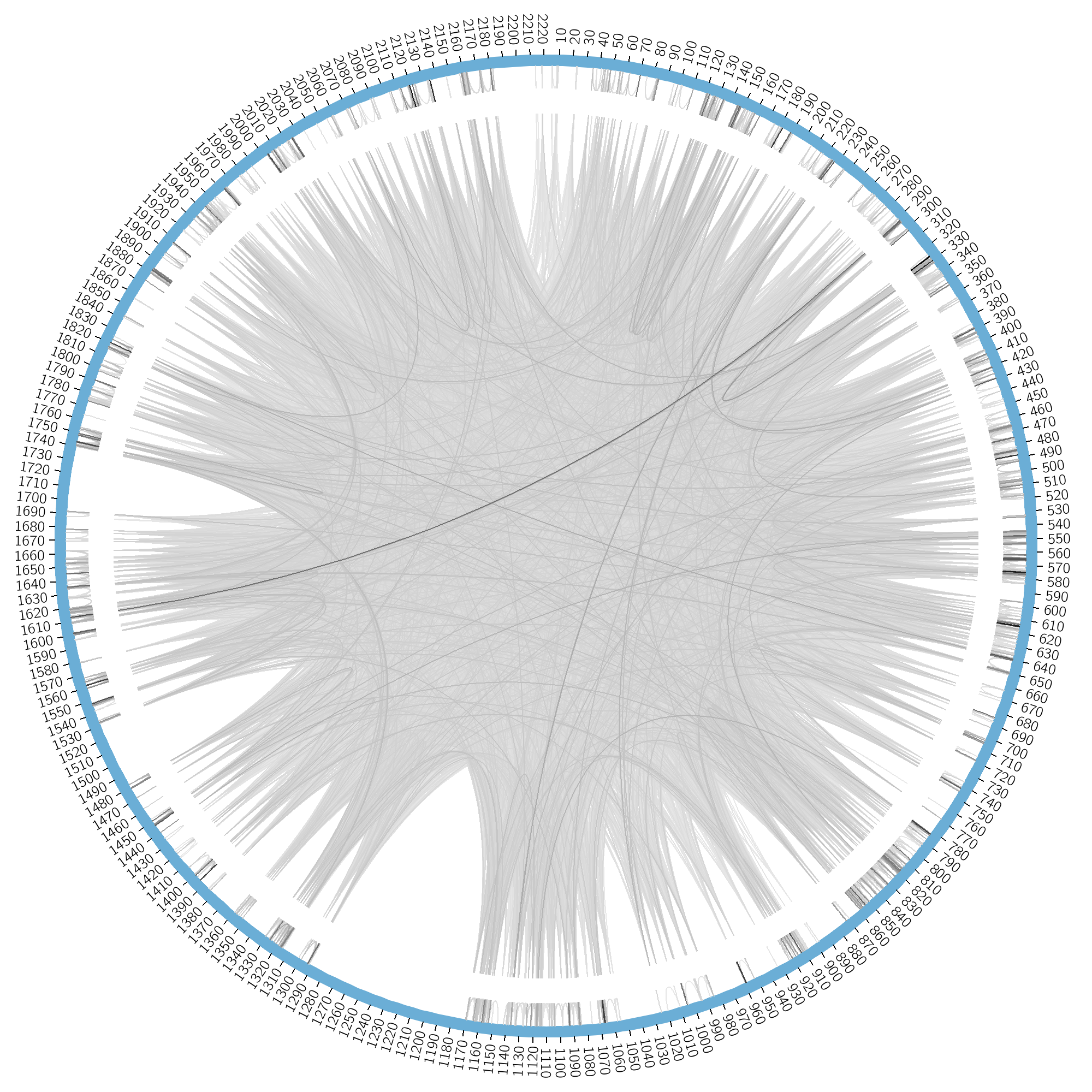}
  \caption{
    \label{fig:ccplm-10000}
    The $10^5$ strongest couplings identified by CC-PLM while retaining $10,000$ largest correlations. The number of loci involved is $5132$. Darkness of edges represent the strength of couplings. The visual impression is quite similar to Figs.~\protect\ref{fig:ccplm-30000} and \protect\ref{fig:ccplm-30000b} above.
  }
\end{figure}

To demonstrate robustness of the ``evaluation-by-visualization''
we show in Figs.~\ref{fig:ccplm-30000b}
and~\ref{fig:ccplm-10000} all the 
$10^5$ strongest couplings obtained in the same procedure
as in Fig.~\ref{fig:ccplm-30000}, and
the results of CC-PLM starting from only 
the loci involved in the $10,000$ largest correlations. 
In both cases the 
long-range inferred couplings are very similar to Fig.~\ref{fig:ccplm-30000}.
Short-range couplings are additionally displayed around the outer rim.

\end{document}